# A gradient-based approach to fast and accurate head motion compensation in cone-beam CT

Mareike Thies, Fabian Wagner, Noah Maul, Haijun Yu, Manuela Goldmann, Linda-Sophie Schneider, Mingxuan Gu, Siyuan Mei, Lukas Folle, Alexander Preuhs, Michael Manhart, Andreas Maier, *Member, IEEE*

*Abstract*—Cone-beam computed tomography (CBCT) systems, with their flexibility, present a promising avenue for direct point-of-care medical imaging, particularly in critical scenarios such as acute stroke assessment. However, the integration of CBCT into clinical workflows faces challenges, primarily linked to long scan duration resulting in patient motion during scanning and leading to image quality degradation in the reconstructed volumes. This paper introduces a novel approach to CBCT motion estimation using a gradient-based optimization algorithm, which leverages generalized derivatives of the backprojection operator for cone-beam CT geometries. Building on that, a fully differentiable target function is formulated which grades the quality of the current motion estimate in reconstruction space. We drastically accelerate motion estimation yielding a 19-fold speed-up compared to existing methods. Additionally, we investigate the architecture of networks used for quality metric regression and propose predicting voxel-wise quality maps, favoring autoencoder-like architectures over contracting ones. This modification improves gradient flow, leading to more accurate motion estimation. The presented method is evaluated through realistic experiments on head anatomy. It achieves a reduction in reprojection error from an initial average of 3 mm to 0.61 mm after motion compensation and consistently demonstrates superior performance compared to existing approaches. The analytic Jacobian for the backprojection operation, which is at the core of the proposed method, is made publicly available. In summary, this paper contributes to the advancement of CBCT integration into clinical workflows by proposing a robust motion estimation approach that enhances efficiency and accuracy, addressing critical challenges in time-sensitive scenarios.

*Index Terms*—computed tomography, deep learning, differentiable programming, gradient descent, motion compensation

## I. INTRODUCTION

CONE-beam computed tomography (CBCT) systems are CT scanning devices that come with a decisive flexibility advantage over helical multi-detector computed tomography (MDCT) scanners. MDCT systems are stationary and patients need to be transported from their point-of-care to the imaging device which might be located in a different area of the hospital [1]. This transport can be problematic for critically-ill patients because it carries the risk of adverse effects in up to 70% of the cases [2]. Furthermore, the transport of the patient to the imaging device takes time, and medical staff accompanying the patient is required [3]. For that reason, specialized and portable CBCT has been considered as an alternative for direct point-of-care imaging, e.g., of the head [3], [4]. Also, patients coming to the hospital with acute stroke symptoms require head imaging. In that case, any delay caused by transporting the patient to the MDCT scanner can have severe negative effects on the patient outcome. For patients with acute ischemic stroke, it is crucial to initiate an endovascular treatment as soon as possible. Upon 150 minutes after symptom onset, their chance to overcome the stroke without any functional impairment drops by 10% to 20% per additional hour [5]. Bringing the patient directly to the angio suite and performing a CBCT scan where the endovascular procedure can be initiated without delay can potentially speed up the workflow substantially and improve patient outcome [1].

Despite these advantages, the integration of CBCT systems into the clinical workflow for stroke or bedside imaging in the ICU is challenging. One primary reason for that is patient motion which can lead to severe deterioration of the reconstructed image. As an average CBCT scan takes from 4 to 30 seconds [6] compared to approximately 0.4 seconds for one gantry rotation in modern MDCT scanners [7], involuntary motion has a more prominent influence on CBCT reconstructions [8]. Cancelliere et al. showed that out of 310 patients presenting at the hospital with acute stroke symptoms and receiving a non-contrast CBCT in the angio suite, 51% exhibited motion artifacts in the reconstructed volume and 11% could not be clinically interpreted due to the severity of the deteriorations in the image [9]. For a successful integration of CBCT systems into the clinical workflow for critically-ill or stroke patients, a reliable method for preventing motion artifacts is therefore desired.

The reduction of overall scan time in modern CBCT systems is beneficial in terms of motion artifacts [10]. However, it is often insufficient to fully suppress involuntary motion. As a result, software-based solutions which estimate the motion patterns from the measured data and compensate for the misalignment during the reconstruction have been proposed. These approaches share a two-step pipeline consisting of (1) motion estimation and (2) a motion-compensated reconstruc-

M. Thies, F. Wagner, N. Maul, H. Yu, M. Goldmann, L. S. Schneider, M. Gu, S. Mei, L. Folle, and A. Maier are with the Pattern Recognition Lab, Friedrich-Alexander-Universität Erlangen-Nürnberg, Erlangen, Germany. N. Maul, M. Goldmann, A. Preuhs, and M. Manhart are with Siemens Healthineers AG, Forchheim, Germany. Corresponding author e-mail: mareike.thies@fau.de.

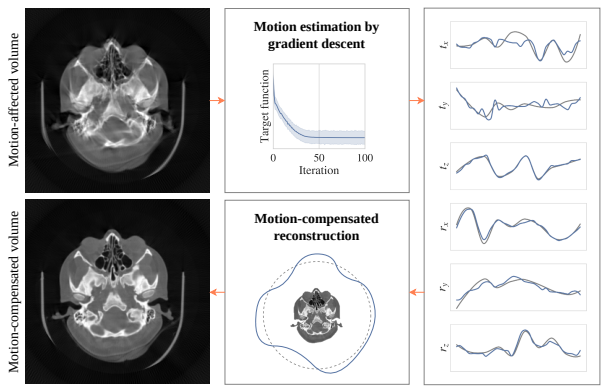

Fig. 1: Rigid motion compensation is performed by first solving an optimization problem to estimate the motion patterns from the measured data. Then, a motion-compensated reconstruction is performed based on the estimated motion patterns to compute the compensated volume.

tion based on the estimated motion patterns from the previous step. Fig. 1 illustrates this two step approach.

Whereas the motion-compensated CT reconstruction is relatively straight-forward given a known motion pattern, the estimation of the scan- and patient-specific motion is the crucial step. To solve this in an image-based manner, the 3D patient motion must be derived from the 2D projection images. This is a non-trivial inverse problem since it depends on both the measured projection data as well as the scanning geometry. Consequently, instead of a direct estimation, it is usually formulated as an optimization problem. A number of target functions have been proposed in prior work. First, these include consistency-based functions which are evaluated on the sinogram and require no intermediate reconstruction [11]–[14]. Second, target functions incorporating the forward model of X-ray imaging have been proposed which maximize the alignment of a usually known, motion-free volume or a simultaneously estimated volume with the measured projection data in sinogram domain in a 2D/3D-registration setting [15]–[17]. Third, target functions have been proposed which measure the image quality on the reconstructed volume itself. These functions incorporate an inverse model of the CT imaging process in each evaluation and are often referred to as autofocus objectives [6], [18]–[21].

Autofocus-type objectives have some decisive advantages over other types of target functions. They can handle truncated data as opposed to consistency-based objectives and do not require a prior motion-free scan in contrast to most forward model-based objectives. However, they also crucially depend on the choice of quality metric quantifying the severity of motion artifacts in the volume. Formulating a discriminative quality metric is challenging since it is a reference-free task meaning that it depends only the potentially motion-corrupted volume. Typical choices therefore rely on basic properties of motion-free images such as sharp edges, piece-wise constancy, or low entropy [21], [22]. Nowadays, research focuses on data-driven quality metrics which are parameterized by deep neural networks [6], [20], [23].

Multiple recent works propose learning-based functions which are shown to regress a certain quality metric robustly on their target anatomy. Still, this type of motion compensation has not prevailed in practice. We identify two main reasons for this: (1) The optimization is too slow. Each evaluation of the target function is expensive and many intermediate reconstructions are needed to converge. (2) Some motion artifacts remain even after optimization. The optimization does not estimate the underlying motion patterns precisely enough.

In this work, we propose a new motion estimation approach for rigid, inter-frame motion in CBCT which addresses both short-comings mentioned above. First, we observe that most existing methods for autofocus motion compensation use gradient-free optimization algorithms which results in high numbers of target function evaluations and long computation times. We propose to replace the optimizer by a gradient-based one. Only few existing studies made use of gradient information in this context before, e.g., by approximating first and second derivatives numerically [24]. Alternatively, differentiable grid sampling and interpolation has been paired with multiple partial angle reconstructions for gradient computation [25]. Instead, building upon our previous work [26], [27], we generalize the derivation and implementation of analytical derivatives of the reconstructed volume to the geometry parameters from fan-beam to cone-beam geometries. This allows us to formulate a fully differentiable autofocus-type target function for CBCT which expresses the quality of an intermediate reconstruction. Since it is differentiable, the gradient of this target function concerning the motion parameters, a subgroup of the geometry parameters, can be obtained and used for gradient-based optimization, thereby speeding up the motion estimation.

Further, we observe that most existing networks for image quality metric regression use a contracting architecture. The spatial dimensions are gradually reduced until only one scalar value is regressed at the output [6], [20], [23]. However, in the gradient-based case, this structure is not ideal as gradients of the scalar output value with respect to the volume tend to behave unstably and have an unevenly distributed amplitude across the input. Thus, we propose to regress voxel-wise quality maps, thereby promoting autoencoder-like architectures over contracting ones for the image quality metric. Spatially resolved quality maps have recently been proven beneficial for deformable CT motion compensation [28]. We show that this modification also leads to more informative gradient information and improved overall motion estimation in a rigid motion compensation framework.

In summary, the contributions of this paper are:
- The generalization of analytic derivatives for CT reconstruction from fan-beam to cone-beam geometries.
- The application of gradient-based optimization for motion estimation in clinical CBCT applications.
- The regression of voxel-wise quality metrics for improved gradient-flow.

## II. METHODS

### A. Rigid motion compensation for CBCT

In CBCT reconstruction, the measured line integrals from different directions are translated into a spatially resolved map of X-ray attenuation coefficients. This process depends fundamentally on the exact knowledge of the scanning geometry describing the 3D relationship between X-ray source, flat-panel detector, and patient. Usually, stillness of the patient during the scan is assumed. In that case, any reproducible, circular scanning trajectory of the CBCT scanner is uniquely characterized by a set of calibrated projection matrices [29] which map a point in the 3D world coordinate system onto a point in the detector coordinate system of a specific projection view. Given accurate projection matrices corresponding to the projection data, there exist a number of analytic and algebraic algorithms to solve the reconstruction problem [30]. Movement of the patient introduces a mismatch between the calibrated projection matrices and the projection data. During reconstruction, this mismatch leads to artifacts in the final image.

The characteristics of motion patterns that can occur in CBCT differ depending on the anatomy. While some regions of the human body tend to move non-rigidly, we assume rigid motion patterns for the head which is non-deformable. In that case, the motion patterns can be parameterized by three rotational and three translational components. Moreover, we require that the frequency of the motion is low compared to the exposure time for a single projection. This justifies the assumption that motion occurs only in between projections (inter-frame) and not during the acquisition of one projection (intra-frame). This is a valid assumption given typical exposure times of around $5 \, \mathrm{ms}$ per projection which is also in accordance with prior work [6], [13], [31].

We aim to eliminate the mismatch between projection matrices and projection data introduced by rigid, inter-frame patient motion by updating the projection matrices. This is drawing from the observation that any rigid motion of the patient inside the scanner can identically be modeled by the inverse rigid transformation applied to the scanner itself. Hence, for each projection $j$, we seek to estimate rigid transformation matrices $\mathbf{T}_j \in \mathbb{R}^{4 \times 4}$ which are multiplied to the initial projection matrices $\mathbf{P}_j \in \mathbb{R}^{3 \times 4}$ to yield updated projection matrices $\hat{\mathbf{P}}_j \in \mathbb{R}^{3 \times 4}$

$$\hat{\mathbf{P}}_j = \mathbf{P}_j \cdot \mathbf{T}_j(\mathbf{x}) \ . \tag{1}$$

The transformation matrices $\mathbf{T}_j$ are parameterized by a vector of free parameters $\mathbf{x} \in \mathbb{R}^N$, where $N$ is the number of free parameters (see section II-B.1). The corrected projection matrices are then reconstructed together with the original projection data to yield the motion corrected reconstruction. The crucial part of this method is the estimation of the transformation matrices $\mathbf{T}_j$ which depend on the projection data and are scan-specific. Here, we solve an unconstrained optimization problem of the form $\mathbf{x}^* = \arg\min_{\mathbf{x}} f(\mathbf{x})$, where $\mathbf{x}^* \in \mathbb{R}^N$ is the optimized set of free parameters which ideally compensates for the patient motion and yields a consistent set of projection data and projection matrices $\mathbf{P}_j^* = \mathbf{P}_j \cdot \mathbf{T}_j(\mathbf{x}^*)$.

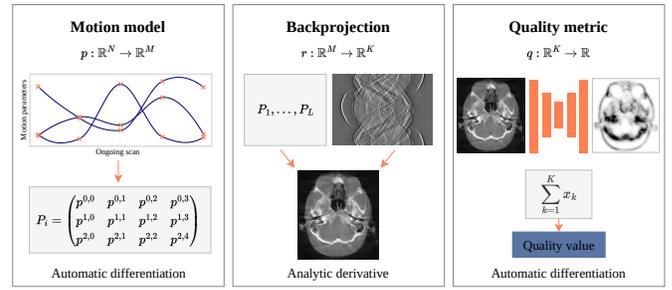

Fig. 2: Overview of the proposed method for rigid motion estimation in CBCT. The three basic building blocks are (1) the motion model, (2) the differentiable backprojection enabling end-to-end gradient flow, and (3) a trained quality metric which regresses spatially resolved quality maps.

In the following, we will detail our proposed method to formulate the target function $f(\mathbf{x})$ and to solve the optimization problem.

### B. Differentiable autofocus pipeline

We propose an autofocus-inspired target function for motion estimation which yields informative gradient updates for the free parameters describing the patient motion. The target function depends only on the projection data and a set of calibrated projection matrices as an initialization of the scanning geometry. The full target function $f(\mathbf{x}) : \mathbb{R}^N \to \mathbb{R}$ consists of three distinct processing steps which are highlighted in Fig. 2. The free parameters $\mathbf{x} \in \mathbb{R}^N$ describing the motion trajectory are processed by the motion model $\mathbf{p} : \mathbb{R}^N \to \mathbb{R}^M$ which computes projection matrices from the initial calibrated matrices and the motion pattern described by $\mathbf{x}$. Here, $M$ is the number of entries in all projection matrices of the scan. Subsequently, the backprojection $\mathbf{r} : \mathbb{R}^M \to \mathbb{R}^K$ uses the updated projection matrices to analytically backproject the filtered projection data into a volume with $K$ elements. Note that, for simplicity, we omit the filtered projection data as an input to the backprojection function $\mathbf{r}$. The backprojection results in an intermediate reconstruction of the data given the current estimate of the free parameters. Eventually, a quality metric $q : \mathbb{R}^K \to \mathbb{R}$ returns a scalar quality value given the intermediate reconstruction. In our proposed method, we first regress a full quality map of the same dimensions as the input volume which is subsequently averaged to a single value. The full target function can be written as $f(\mathbf{x}) = q(\mathbf{r}(\mathbf{p}(\mathbf{x})))$. To compute the gradient of the target function to the free parameters, we apply the chain rule

$$\frac{\mathrm{d}f}{\mathrm{d}\mathbf{x}} = \frac{\mathrm{d}q}{\mathrm{d}\mathbf{r}} \cdot \frac{\mathrm{d}\mathbf{r}}{\mathrm{d}\mathbf{p}} \cdot \frac{\mathrm{d}\mathbf{p}}{\mathrm{d}\mathbf{x}} \ . \tag{2}$$

Here, $\frac{\mathrm{d}q}{\mathrm{d}\mathbf{r}} \in \mathbb{R}^{1 \times K}$ is the gradient of the quality metric to the intermediate reconstruction. $\frac{\mathrm{d}\mathbf{r}}{\mathrm{d}\mathbf{p}} \in \mathbb{R}^{K \times M}$ is the Jacobian of the backprojection function and $\frac{\mathrm{d}\mathbf{p}}{\mathrm{d}\mathbf{x}} \in \mathbb{R}^{M \times N}$ is the Jacobian of the motion model. In the following sections II-B.1, II-B.2, and II-B.3, we will explain each of the three steps in detail and present how the corresponding derivative can be obtained. Furthermore, in section II-B.3, we will highlight why the regression of a spatially resolved quality map followed by an average operator is superior in the gradient-based setting.

*1) Motion model:* In general, the projection geometry of any CBCT scan is uniquely defined by a set of projection matrices $\mathbf{P}_j$, $j = 1, \ldots, N_p$, where $N_p$ is the number of projections in a scan. This is the most flexible form to describe an unconstrained CBCT trajectory. Each projection matrix is only unique up to scale and not every $3 \times 4$ matrix is a valid projection matrix. Therefore, it is not advisable to directly optimize for the projection matrix entries. Instead, for rigid motion compensation, prior knowledge restricts the degrees of freedom to six per view. Without further constraints, the rigid motion state of each projection is unrelated to that of the neighboring views. In a typical CBCT scan, however, multiple projection images are acquired per second. Hence, we impose an additional temporal smoothness constraint on the motion patterns by fitting an Akima spline to each of the six motion parameters $(t_x, t_y, t_z, r_x, r_y, r_z)$ separately. The spline is parameterized by $N_n$ nodes. Given that $N_n < N_p$, this reduces the degrees of freedom from $6N_p$ to $6N_n$ and enforces each motion parameter to vary smoothly throughout the scan. This is a reasonable assumption given that the typical types of head motion were identified to be slow oscillations as well as uniformly increasing deviations from an initial position [8]. Furthermore, our parameterization is in accordance with prior work which relied identically on Akima splines [20], other types of smooth functions [11], [32], or regularization terms in the target function which implicitly punish non-smooth motion patterns [6], [19], [33].

As motion model we refer to the function that translates $N = 6N_n$ parameters, stored in the vector $\mathbf{x}$ and describing Akima spline node values for each motion parameter, into $M = 12N_p$ projection matrix entries. This is achieved by evaluating the splines at the time point of each projection and updating a set of initial projection matrices according to equation 1. Since this is a relatively cheap computation, we implement the splines in native *PyTorch* and obtain the corresponding derivatives automatically. If not stated differently, we utilize ten nodes for the simulation of artificial motion patterns and 30 nodes for the motion patterns being estimated by our method. The nodes are evenly spaced across the temporal direction of the scan with one node at the beginning and one node at the end of the scan range.

*2) Analytic derivative for CBCT backprojection operator:* An analytic backprojection is implemented to perform intermediate reconstructions given the iteratively updated projection matrices returned from the motion model. In addition to the projection matrices, the backprojection depends on an appropriately filtered version of the projection images. Since we consider projection data acquired on a full circular trajectory, we apply a classical shift-invariant ramp filter. During backprojection, the reconstructed value $I$ at position $\mathbf{p}^{\text{world}} = (x, y, z)^\top$ in world coordinates is computed as the sum over the values of the filtered projection data at the forward projected positions in each projection image

$$I(\tilde{\mathbf{p}}^{\text{world}}) = \sum_{j=1}^{N_p} d_j \left( \mathbf{g} \left( \mathbf{s}_j \left( \tilde{\mathbf{p}}^{\text{world}} \right) \right) \right) \ . \tag{3}$$

Here, $N_p$ is the number of projections in the scan and $\tilde{\mathbf{p}}^{\text{world}} = (x, y, z, 1)^\top \in \mathbb{P}^{3+}$ is the homogeneous representation of $\mathbf{p}^{\text{world}}$ following the same notation for homogeneous coordinates as used in Aichert et al. [13]. Function $\mathbf{s}_j : \mathbb{P}^{3+} \to \mathbb{P}^{2+}$ computes the forward projected homogeneous position of $\tilde{\mathbf{p}}^{\text{world}}$ in the detector coordinate system of the $j$-th projection image by multiplication with the $j$-th projection matrix $\mathbf{s}_j(\tilde{\mathbf{p}}) = \mathbf{P}_j \cdot \tilde{\mathbf{p}}$. Subsequently, $\mathbf{g} : \mathbb{P}^{2+} \to \mathbb{R}^2$ restores the euclidean coordinate from the homogeneous coordinate by dividing all elements by the last component of the vector. Finally, $d_j : \mathbb{R}^2 \to \mathbb{R}$ bilinearly interpolates the $j$-th filtered projection image at the forward projected position and returns the corresponding value which is summed up over all $N_p$ projections to yield the final backprojected value $I$.

To reconstruct a full volume, equation 3 needs to be applied to all voxel positions in world coordinates which is extremely time intensive when evaluated sequentially and prohibits the straight-forward use of automatic differentiation. For that reason, we implement a custom GPU-accelerated version of the backprojection and derive and implement the corresponding analytic expression for the derivative. This allows us to have GPU-accelerated implementations for backprojection and gradient computation and still connect the reconstructed image and the projection matrices in a differentiable manner. In our previous paper, we have proposed a similar differentiable formulation for fan-beam CT backprojection [26]. However, it is not applicable to cone-beam geometries and, therefore, its practical applicability is limited. In the following, we generalize that formulation to cone-beam geometries.

We give an expression for the gradient of the backprojected value at position $\mathbf{p}^{\text{world}} = (x, y, z)^\top$ to the 12 entries of the $\gamma$-th projection matrix $\mathbf{P}_\gamma$. For that, we define $\tilde{\mathbf{p}}_\gamma^{\text{detector}} = \mathbf{s}_\gamma(\tilde{\mathbf{p}}^{\text{world}}) = (u_\gamma, v_\gamma, w_\gamma)^\top \in \mathbb{P}^{2+}$ as the homogeneous forward projected position in detector coordinates. The gradient $\mathbf{G}_\gamma \in \mathbb{R}^{3 \times 4}$ is

$$\mathbf{G}_\gamma = \begin{pmatrix} \frac{g_{u,\gamma} x}{w_\gamma} & \frac{g_{u,\gamma} y}{w_\gamma} & \frac{g_{u,\gamma} z}{w_\gamma} & \frac{g_{u,\gamma}}{w_\gamma} \\ \frac{g_{v,\gamma} x}{w_\gamma} & \frac{g_{v,\gamma} y}{w_\gamma} & \frac{g_{v,\gamma} z}{w_\gamma} & \frac{g_{v,\gamma}}{w_\gamma} \\ -\frac{h(x)}{w_\gamma^2} & -\frac{h(y)}{w_\gamma^2} & -\frac{h(z)}{w_\gamma^2} & -\frac{h(1)}{w_\gamma^2} \end{pmatrix} \ , \tag{4}$$

with

$$h(i) = g_{u,\gamma} i u_\gamma + g_{v,\gamma} i v_\gamma \ . \tag{5}$$

To obtain the values $g_{u,\gamma} \in \mathbb{R}$ and $g_{v,\gamma} \in \mathbb{R}$, the spatial derivative of the filtered projection data along the two axes of the detector coordinate system is computed. It is numerically approximated by second order accurate central differences along the horizontal and vertical axis of each filtered projection image yielding two gradient images per view. These are interpolated bilinearly at the euclidean detector position $(\frac{u_\gamma}{w_\gamma}, \frac{v_\gamma}{w_\gamma})^\top \in \mathbb{R}^2$. The interpolated values from the $\gamma$-th gradient image with respect to the first and second detector axis are denoted $g_{u,\gamma}$ and $g_{v,\gamma}$, respectively. All investigations are restricted to the discretized problem setting, where projection images are measured at discrete angles and in a discrete detector coordinate system. We refer to our previous work for a more in-depth comparison between the continuous derivation and the corresponding discretized version [26].

Equation 4 provides the partial derivatives of the back-projected value at one position $\mathbf{p}^{\text{world}}$ in the volume to the twelve entries of one projection matrix $P_\gamma$. It needs to be evaluated for all $K$ discrete volume positions as well as for all $N_p = M/12$ projection matrices to fill the full Jacobian $\frac{d\mathbf{r}}{d\mathbf{p}}$. The implementation of GPU-accelerated backprojection and corresponding gradient computation is openly available at https://github.com/mareikethies/geometry_gradients_CT. All computations are wrapped in a custom *PyTorch* function which is directly compatible with the automatic differentiation functionality of the library. Due to memory constraints, we never explicitly evaluate the full Jacobian, but directly compute Jacobian-vector products given the gradient signal of the downstream operations.

*3) Pixel-wise quality metric regression:* Given an intermediate reconstruction from the backprojection, we seek to quantify its quality in terms of motion artifacts. At inference time, this mapping needs to be reference-free, meaning that the quality is estimated given only a motion-affected volume and no motion-free reference. To solve this task, a neural network is trained on a simulated paired data set of motion-affected and corresponding motion-free scans to approximate a reference-based target metric given the motion-affected volume only. This has been proposed similarly in prior work [6], [20], [23]. In contrast to existing approaches, we are using the trained network in a gradient-based setting where derivatives of the network output to the input volume are computed for motion estimation. For gradient-free methods, an accurate regression of the quality metric from the input volume is the crucial part. In the gradient-based case, an informative volume gradient obtained via backpropagation through the trained network is equally important as the forward mapping. To ensure that an informative volume gradient is obtained, we propose to train a volume-to-volume network which regresses a full spatially resolved quality map. This is in contrast to existing approaches which directly regress a scalar quality metric using contracting architectures. To obtain a scalar for optimization, we then simply average the volumetric quality map, ensuring that each spatial position has equal contribution to the final value.

As the regression target, we choose the visual information fidelity (VIF) [34]. This metric quantifies the information present in the reference image as well as the loss of information that can be attributed to some distortion process, motion in our case. It explicitly models the human visual system and is thought to provide an image similarity measure between distorted and reference image that aligns well with human perception [6]. The original formulation of VIF computes a scalar where 0 is the lower bound which indicates that all information from the reference image has been lost due to the distortion and 1 indicates that the distorted image contains the same information as the reference, i.e., they are identical [34]. Theoretically, the VIF can be greater than 1 in cases where the image under consideration has a higher information content than the reference. To obtain a spatially resolved volumetric map of the VIF, we follow the approach by Shao et al. [35].

Given a paired data set of motion-free and motion-affected head CT volumes, we compute their corresponding VIF map by setting the motion-free volume as reference image and the motion-affected volume as distorted image. The VIF map is further scaled by the number of voxels $K$ such that an average operation yields values in the range $[0, 1]$. As the final regression target for the network we use $\mathbf{VIF}^*(\mathbf{I}_{\text{dist}}, \mathbf{I}_{\text{ref}}) = 1 - K \cdot \mathbf{VIF}(\mathbf{I}_{\text{dist}}, \mathbf{I}_{\text{ref}})$. We then train a 3D U-net architecture [36] to regress the adjusted VIF map $\mathbf{VIF}^*$ from the motion-affected volume. The U-net consists of basic building blocks with 3D convolutions and ReLU activation function. The number of feature maps per level is $8^l$ with $l = 1, \ldots, 4$ levels. Because we train a regression task, the final layer is a $1 \times 1$ convolution and no further final activation function is used. The model is trained with an L1-loss and Adam optimizer with a learning rate of 0.001 and a batch size of 16 with input and output volumes of size $128 \times 128 \times 128$. After training, the predicted map is simply averaged to get a scalar value. The trained network performs a reference-free mapping and does not require the motion-free volume anymore.

### C. Gradient-based optimization algorithm

The average of the predicted quality map is minimized to obtain a scan specific motion estimate $\mathbf{x}^*$. Importantly, at optimization time, the parameters of the quality metric network are fully trained and frozen. The majority of existing motion compensation algorithms utilize a gradient-free optimizer to minimize the objective with respect to the motion parameters. For gradient-free optimization algorithms it suffices to specify the target function alone whereas gradient-based algorithms require an expression of the gradient which can be hard to formulate because the dependency of the objective value on the motion parameters is not trivial [37]. We are the first to use a fully differentiable autofocus target function which allows us to apply gradient-based optimization. Specifically, we utilize a basic gradient descent update scheme of the form

$$\mathbf{x}^{(n+1)} = \mathbf{x}^{(n)} - s^{(n)} \cdot \frac{df}{d\mathbf{x}} \ . \quad (6)$$

Updated estimates $\mathbf{x}^{(n)}$ are iteratively computed from an initial estimate $\mathbf{x}^{(0)} = \mathbf{0}$ by taking a step into the negative gradient direction scaled by the step size $s^{(n)} \in \mathbb{R}$. The step size is subject to an exponential decay such that $s^{(n)} = s_0 \cdot t^n$ with $t \in \mathbb{R}$ being the decay factor and $s_0 \in \mathbb{R}$ describing the initial step size. If not stated differently, we run the algorithm for 100 iterations with an initial step size of $s_0 = 100$ and a decay factor of $t = 0.97$.

## III. DATA

The CQ500 data set is used for our experiments [38]. Originally, this data set includes 491 reconstructed head CT scans acquired on different MDCT scanners. We first filter the data set for scans which have been reconstructed with a small slice thickness. From the remaining scans we further exclude those which have considerably fewer or more slices than the average sample. This results in 320 scans which we split sequentially on patient level into training set (150), validation set (50), and test set (120). After filtering, all scans have a reconstructed slice thickness of $0.625\,\text{mm}$ and an isotropic in-slice spacing between $0.38\,\mathrm{mm}$ and $0.58\,\mathrm{mm}$ (mean of $0.472\,\mathrm{mm}$). Because we are working on cone-beam geometry, we simulate the corresponding detector data from the reconstructed MDCT volumes. Each volume is forward projected on a circular trajectory in cone-beam geometry using the software *PyroNN* [39] with 360 projections on a full $2\pi$ angular range, a source-to-isocenter distance of $785\,\mathrm{mm}$, a source-to-detector distance of $1200\,\mathrm{mm}$, and a detector of shape $500 \times 700$ pixels of isotropic pixel spacing ($0.64\,\mathrm{mm}$). A ramp filter and a cosine filter are applied to the computed projection images.

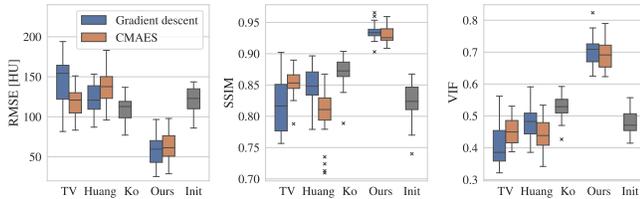

Fig. 3: Comparison of the proposed quality metric with total variation (TV) and the network-based quality metric proposed by Huang et al. [6] as well as the image-based one-step method by Ko et al. [40]. Metrics are root mean squared error (RMSE) (↓), SSIM (↑), and VIF (↑) which are computed on the motion-compensated reconstructed volumes.

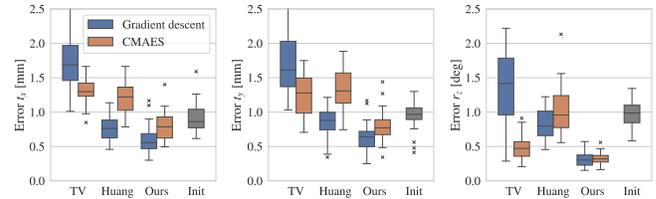

(a) In-plane motion parameters

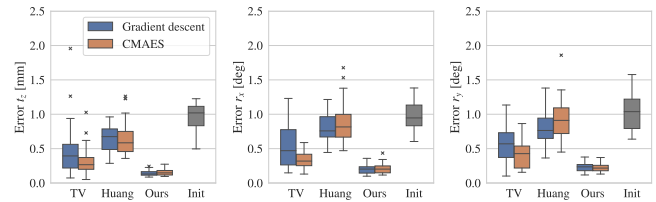

(b) Out-of-plane motion parameters

Fig. 4: Comparison of the proposed quality metric with total variation (TV) and the network-based quality metric proposed by Huang et al. [6]. The upper (4a) and lower (4b) plots depict the mean absolute error (MAE) (↓) for the in-plane motion parameters ($t_x$, $t_y$, and $r_z$) and out-of-plane motion parameters ($t_z$, $r_x$, and $r_y$), respectively, referring to the plane in which the source rotates.

To train the quality metric network (section II-B.3), we dynamically sample a new random motion perturbation each time a sample is used for training and apply it to the projection matrices. The spline-based motion model (section II-B.1) with 10 nodes per spline is used with a maximal amplitude of $10\,\mathrm{mm}$ for translation parameters and $15°$ for rotation parameters. To ensure that all motion states that could be encountered during optimization are represented in the training data, we include motion patterns with unequal amplitude across the different motion parameters as well as motion patterns that perturb the data only slightly (see supplementary material for pseudo code). All splines are individually zero-centered and translated into perturbed projection matrices using the motion model. The filtered projection data is reconstructed from these perturbed matrices and the training target $\mathbf{VIF}^*$ is computed from the perturbed and unperturbed reconstruction as described in section II-B.3. The intensities of the reconstructed volumes are normalized with fixed, sample-independent offset and slope to lie approximately in the interval $[0, 1]$ before feeding them to the network.

## IV. EXPERIMENTS AND RESULTS

All motion compensation experiments are performed on the same 30 patients from the test set for quality metric model training. A random motion pattern is sampled for each patient with an amplitude of $5\,\mathrm{mm}$ for translation and $5°$ for rotation which is kept constant across different methods and optimization algorithms. The motion estimation itself runs on grids of size $128 \times 128 \times 128$ with a $2\,\mathrm{mm}$ spacing for the reconstructed signal, but the image-based results are computed on higher resolved signals with $256 \times 256 \times 256$ voxels of size $1\,\mathrm{mm}$ for the final motion-compensated reconstruction. For the evaluation we rigidly register all motion-compensated reconstructions to their respective ground truth reconstruction in 3D. All box plots in this paper show the median and inter-quartile range as well as the minimum and maximum values. Outliers are highlighted by cross markers. Values of the initial metrics before motion compensation are shown by the gray box. Additionally, we apply our method to two clinical cone-beam CT head scans that are affected by real patient motion. Since the motion patterns underlying these scans are introduced by the patients during the scanning procedure, a quantitative analysis is not possible in this case, but qualitative results are presented in section IV-D.

### A. Motion compensation performance

To assess the motion compensation performance of our proposed method, we compare it to three previously published reference methods, two of which are optimization-based algorithms. For a fair comparison, the data setup, motion model, and backprojection step is identical in all methods, but only the quality metric is replaced. First, 3D total variation (TV) is chosen as a classical autofocus quality metric which enforces sparse volume gradients via minimization of the L1-norm. It is a parameter-free quality metric and has been shown to yield informative information for image alignment [20], [22]. Second, we re-implement the quality metric network proposed by Huang et al., which also regresses the VIF measure using a 3D architecture [6]. In contrast to our quality metric, it is not trained on spatially resolved quality maps, but the authors use a contracting residual convolutional architecture followed by a fully-connected layer for the final regression task. We train the network on the same data as our proposed network, but directly compute the spatial average of the VIF as the training target. The learning rate is set to $0.0001$ and the weights are optimized by the Adam algorithm for $500$ epochs. Third, we compare our method to the approach proposed by Ko et al. [40] which is a pure image-to-image approach mapping motion-affected input volume to motion-corrected output volume. In contrast to our approach and the optimization-based reference methods, it does not estimate underlying motion patterns explicitly. We use the code released by Ko et al. at https://github.com/youngjun-ko/ct_mar_attention and prepare the training data in the same way as for all other methods.

As can be seen in Fig. 3, our method consistently achieves the best results concerning metrics computed on the final motion-compensated volume. Focusing on the blue box plots for gradient-based optimization, our proposed method consistently improves upon the initial motion-affected reconstruction (gray box) in terms of root mean squared error (RMSE), SSIM, and VIF, reaching an SSIM of more than $0.90$ for all cases in the test set. In comparison, both other optimization-based methods (TV target function and the method by Huang et al.) cannot stably improve upon the initial state and the performance of the image-to-image approach by Ko et al. also achieves inferior results. The reason for this observation can be identified in Fig. 4. While all three optimization-based methods successfully improve upon the initial motion-affected state in terms of out-of-plane parameters, this is not the case for the in-plane parameters. Optimization using TV as quality metric results in particularly large in-plane errors, often larger than in the initial state while leading to accurate fits for the out-of-plane parameters. The method by Huang et al. yields improved in-plane and out-of-plane parameters, but the improvement is marginal in some cases. For our method, too, out-of-plane motion is recovered almost perfectly with absolute errors below $0.5\,\text{mm}$ and $0.5°$ in all cases, whereas in-plane parameters $t_x$ and $t_y$ exhibit a median absolute error of more than $0.5\,\text{mm}$. In Fig. 5, we analyze the reprojection error (RPE), which is computed by forward projecting a fixed set of 300 3D points with radii $25\,\text{mm}$, $50\,\text{mm}$, and $100\,\text{mm}$ around the isocenter onto the detector planes using the recovered and target geometries. Our method outperforms both comparison methods by a large margin improving from an initial median RPE of around $3\,\text{mm}$ to RPEs below $1\,\text{mm}$ in all cases. Table I summarizes the mean of all metrics for the gradient-based case.

Fig. 6 validates our findings by two qualitative examples. The reconstructed slices confirm that our method successfully compensates motion artifacts. The output of our proposed algorithm is visually close to the ground truth. The method proposed by Huang et al. can reduce motion artifacts, but visible streaks and blurred edges remain, especially for example 2. For the TV quality metric, we can visually confirm the previous result that it leads to acceptable results for out-of-plane motion (mostly visible in the sagittal views), but fails to recover in-plane motion (see axial views). The algorithm by Ko et al. produces visually very clean images. However, the anatomy appears altered in shape and structure at multiple locations such as the upper part of the head in the axial view of example 2. A small bone structure can be viewed in the enlarged regions of interest (ROI) for each slice. Only our method can fully recover the shape and structure of all details of these ROIs and successfully removes streaks and blurred edges.

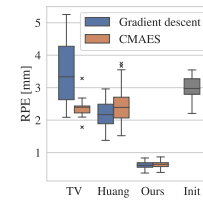

Fig. 5: Comparison of the proposed quality metric with total variation (TV) and the network-based quality metric proposed by Huang et al. [6]. The plot shows the reprojection error (RPE) (↓) which quantifies the distance between points after forward projection onto the detector planes thereby emphasizing on measurable deviations.

Fig. 7 visualizes the motion patterns across the scan range for the first qualitative example in Fig. 6. The motion curves estimated by our proposed method are closest to the ground-truth motion curves, but with a slightly higher error for in-plane translations $t_x$ and $t_y$ than for the other four parameters. The reference methods exhibit larger deviations from the ground truth with particularly large errors for the in-plane motion estimated with TV quality metric. These findings correlate well with the characteristics of the reconstructed images in Fig. 6.

### B. Optimizer comparison

To analyze the impact of the gradient-based optimization, we compare our method as well as the two existing iterative approaches in a gradient-based and in a gradient-free scenario. For the gradient-free setting, we choose the covariance matrix adaptation evolution strategy (CMAES) algorithm [41] because it is frequently applied for similar problems in related work [19], [23]. It is configured with an initial standard deviation of $0.5\,\text{mm}$ and $0.5°$ for translations and rotations, respectively, and iterated until the number of target function evaluations reaches 10000.

Fig. 8 presents the average runtime required in both optimization settings. Whereas all three investigated methods require less than two minutes when using the proposed gradient-based updates, the same optimization takes substantially longer (more than 30 minutes) when using CMAES. It can be observed that the different quality metrics vary slightly in the time needed for evaluation and gradient computation, but this effect is negligible when compared against the difference between the two optimization strategies. For example, for our proposed quality metric, the CMAES optimization is approximately 19 times slower than the gradient descent algorithm. All runs have been executed using the same implementation for all processing steps except the quality metric and on the same hardware (Nvidia A40 GPU). Moreover, from figures 3, 4, and 5, we can observe that the choice of optimization algorithm does not influence the performance of the motion estimation for our proposed method. All metrics are at least the same (RPE or MAE for out-of-plane motion parameters) or even slightly better (image-based metrics or MAE for in-plane motion parameters) when using gradient descent compared to CMAES for our quality metric. We conclude, that the gradient-based optimization speeds up all three investigated methods by a factor of approximately 19. At the same time, the performance of the motion estimation is either not affected or even slightly improved for both quality metrics which are parameterized by neural networks.

| | RMSE ($\downarrow$) | SSIM ($\uparrow$) | VIF ($\uparrow$) | RPE ($\downarrow$) | MAE $t_x$ ($\downarrow$) | MAE $t_y$ ($\downarrow$) | MAE $t_z$ ($\downarrow$) | MAE $r_x$ ($\downarrow$) | MAE $r_y$ ($\downarrow$) | MAE $r_z$ ($\downarrow$) |
|---|---|---|---|---|---|---|---|---|---|---|
| Init | 120.75 | 0.83 | 0.48 | 3.00 | 0.92 | 0.94 | 0.97 | 0.97 | 1.03 | 0.96 |
| TV | 144.97 | 0.82 | 0.41 | 3.44 | 1.76 | 1.73 | 0.46 | 0.53 | 0.61 | 1.39 |
| Huang et al. | 122.63 | 0.85 | 0.48 | 2.19 | 0.77 | 0.87 | 0.65 | 0.80 | 0.81 | 0.83 |
| Ko et al. | 108.67 | 0.87 | 0.53 | - | - | - | - | - | - | - |
| Ours | 58.49 | 0.94 | 0.70 | 0.61 | 0.61 | 0.64 | 0.14 | 0.20 | 0.23 | 0.32 |

TABLE I: Average quantitative values for all investigated metrics and motion compensation methods optimized with gradient descent. Only image-based metrics can be computed for the method by Ko et al. since it does not explicitly estimate motion patterns.

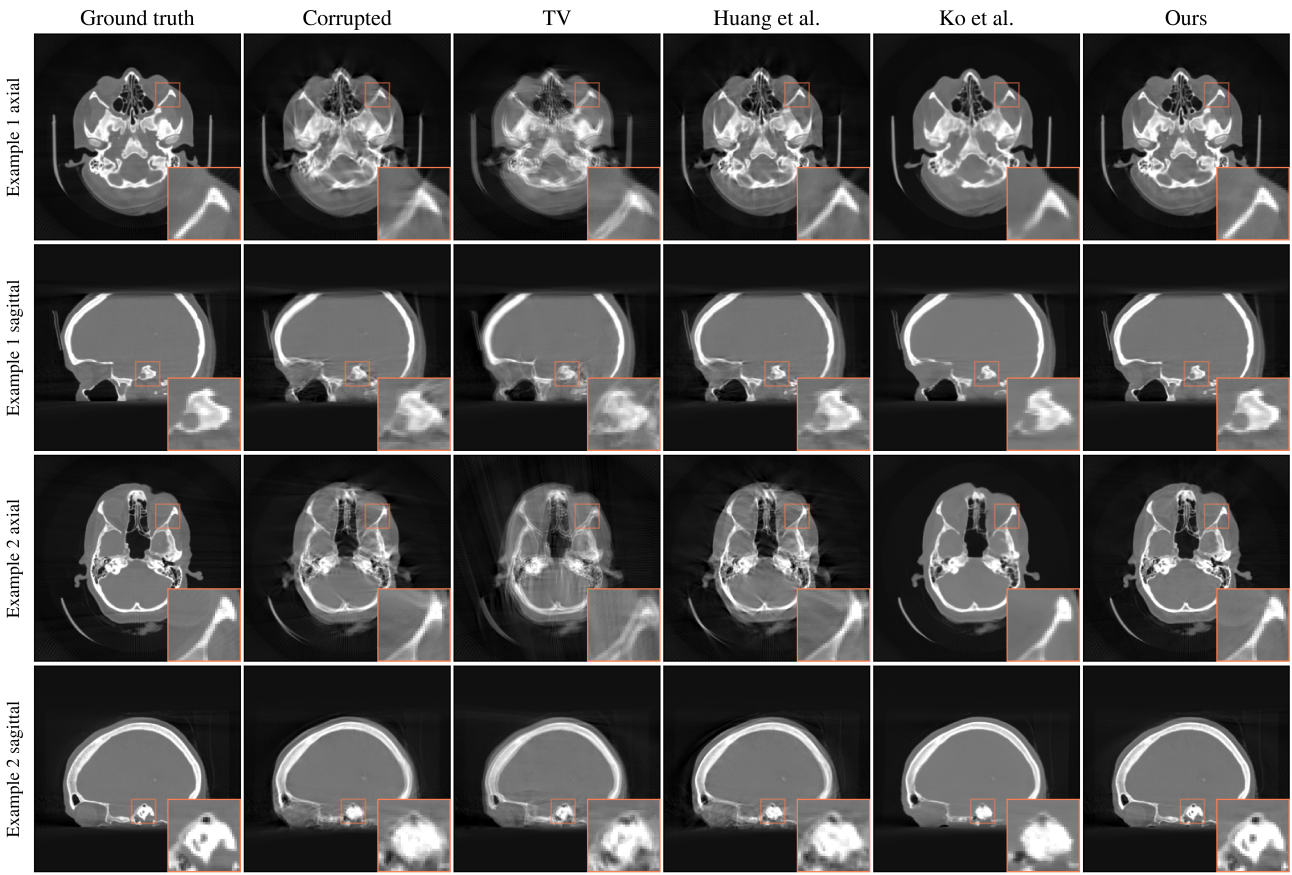

Fig. 6: Qualitative reconstructions for two example patients. For each patient, the upper row shows an axial slice and the lower row shows a sagittal slice through the volume. A $\times 3$ zoom of a region of interest is inserted. All gray values are windowed between $-1200\,\mathrm{HU}$ and $1500\,\mathrm{HU}$.

### C. Quality metric choice

In our proposed method, we use a voxelized VIF map as regression target for the trained quality metric. Recent work identified VIF to be a suitable metric for motion compensation [6]. Nevertheless, any metric which can be computed in a per-voxel manner is a valid choice for our algorithm. Therefore, we evaluate the influence of the target metric on the performance of our method by training the quality metric to regress the mean squared error (MSE) as well as the structural similarity index measure (SSIM) from the motion-affected input. For SSIM, similar to the strategy described for VIF, we regress an adjusted map ensuring that an average value of 0 corresponds to the ideal image via $\mathbf{SSIM}^*(\mathbf{I}_{\text{dist}}, \mathbf{I}_{\text{ref}}) = 1 - \mathbf{SSIM}(\mathbf{I}_{\text{dist}}, \mathbf{I}_{\text{ref}})$. The training setup is not altered except for the computation of the regression target.

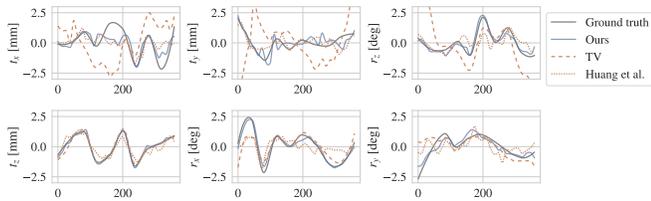

Fig. 7: Motion patterns belonging to example 1 in Fig. 6 plotted as a function of acquired projections. The upper and lower row depicts the motion curves for the in-plane motion parameters ($t_x$, $t_y$, and $r_z$) and out-of-plane motion parameters ($t_z$, $r_x$, and $r_y$), respectively, referring to the plane in which the source rotates. The perturbing ground-truth motion pattern is parameterized by ten spline nodes, whereas all recovering motion patterns are parameterized by 30 nodes. The initial motion-affected reconstruction in Fig. 6 corresponds to motion curves initialized by zeros.

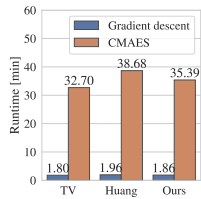

Fig. 8: Comparison of the runtime for the proposed method and the alternative quality metrics in a gradient-based and gradient-free optimization setting. Each bar depicts the average runtime across all test samples for motion estimation excluding the final full-resolution motion-compensated reconstruction.

We observe that all three networks trained to predict quality maps for MSE, SSIM, and VIF, respectively, result in accurate motion estimation and outperform the comparison methods in terms of RPE. VIF achieves the smallest RPE with a mean of $0.61\,\text{mm}$, whereas SSIM and MSE both result in a mean RPE of $0.98\,\text{mm}$. As a result, VIF works most accurately in the given experimental setting, but only with a small margin compared to MSE and SSIM.

### D. Clinical validation

For clinical validation, we apply our method to two real cone-beam CT scans which are noticeably affected by real patient motion (see Fig. 9, left column). The scans were recorded on an Artis Q C-arm interventional angiography system (Siemens Healthineers AG, Forchheim, Germany) with 496 projection images on a $200°$ short scan trajectory. Projection images were acquired with an image size of $960 \times 1240$ pixels, an effective pixel size of $0.316\,\text{mm} \times 0.316\,\text{mm}$ after $2 \times 2$ binning, preprocessed by a vendor-specific algorithm including conversion to line integral domain and a scatter correction, and down-sampled by a factor of four in both dimension for our experiments. We further apply cosine weights, Parker weights, and the ramp filter. Truncation correction is implemented by extending the sinograms by 10% of their width on either side and linearly extrapolating the left and right gray values to zero. The source-to-isocenter distance is $750\,\text{mm}$ and the source-to-detector distance is $1245\,\text{mm}$. Reconstructions are computed identically to the other experiments on a grid of size $128 \times 128 \times 128$ with a $2\,\text{mm}$ spacing for motion compensation and on $256 \times 256 \times 256$ pixels with $1\,\text{mm}$ spacing for visualization. Importantly, we do not retrain or change the network for quality metric regression. Compared to the other experiments, we adjust the step size for the gradient descent to $s_0 = 10$ and we apply a cylindrical mask to the reconstructions before feeding them to the quality network. Next to the proposed algorithm, we further apply the methods by Huang et al. and Ko et al. to the real data. Optimization with TV is excluded since it already produced unsatisfactory results in the previous experiments.

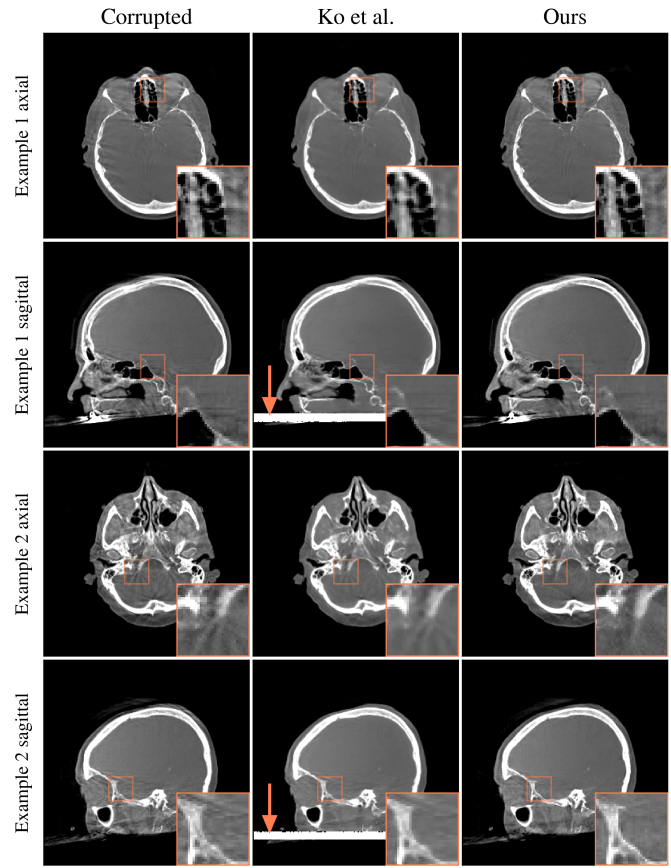

Fig. 9: Qualitative reconstructions for two example patients using real motion-affected clinical cone-beam CT scans. For each patient, the upper row shows an axial slice and the lower row shows a sagittal slice through the volume. A $\times 3$ zoom of a region of interest is inserted.

In Fig. 9, both clinical examples generally exhibit slightly less pronounced motion artifacts than in the previous experiments using spline-based motion patterns. Nevertheless, the typical streak artifacts are clearly visible. After motion compensation with our proposed method, these streaks are largely reduced as can be seen in the zoomed regions. Furthermore, the fine structures in the axial view of example 1 are more clearly visible after motion compensation. Since the ground-truth underlying motion patterns are unknown in this case, a quantitative analysis is not possible. Instead, we manually track an anatomical landmark across the 496 projection images. We choose a highly absorbing tooth implant for example 1 and the top part of the skull for example 2. In Fig. 10, we plot the vertical pixel position of this landmark in the projection images against the motion curves recovered by our method. For clarity, the motion parameter with the highest maximal amplitude in the recovered signal ($r_x$ for example 1 and $t_z$ for example 2) is highlighted in orange while the other five parameters are plotted in gray. For both real scans, the projected motion pattern of the anatomical landmark is very similar to the pattern of the strongest recovered motion parameter. We do not expect identical signals since we omit the influence of the other five motion parameters as well as the projective nature of the landmark positions. Nevertheless, it confirms the validity of the recovered motion patterns. The recovered patterns agree with the impression obtained from the projection images that the most prominent motion is a nodding motion in the first 100 projection images for example 1 and a periodic translation in $z$ for example 2 potentially caused by heavy breathing of the patient.

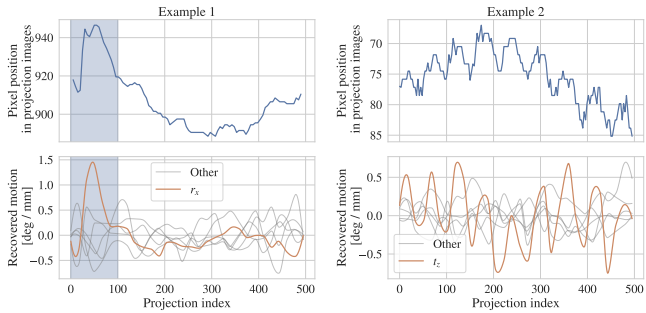

Fig. 10: Comparison of recovered motion patterns for clinical example 1 (left) and example 2 (right) in Fig. 9 with observable motion in projection images. The upper plots describe the vertical pixel position of an anatomical landmark in the projection images extracted by manual annotation. The lower row plots the recovered motion curves (rotations and translations). The motion parameter with the highest amplitude is highlighted in orange ($r_x$ for example 1 and $t_z$ for example 2). Projected motion of the landmark and recovered motion curve with highest amplitude exhibit a strong similarity.

The method by Ko et al. also effectively reduces motion artifacts. However, since it is a pure image-to-image mapping that does not recover the underlying motion pattern, the same analysis as for the proposed algorithm is not possible. Moreover, it predicts an erroneous high-intensity signal for the caudal region of the head (indicated by orange arrows in Fig. 9) which strongly distorts and overlays the actual anatomical structures in that region. The algorithm by Huang et al. performs poorly when transitioning to the real data and its output appears worse than the initial motion-affected image.

## V. DISCUSSION

The experiments demonstrate that by utilizing gradient descent instead of the gradient-free CMAES for optimization, we speed up the motion estimation by a factor of 19. This reduction in runtime is relevant in clinical practice, when motion-compensated reconstructions need to be available within few minutes. For example, in the case of stroke imaging in the angio suite, the time to treatment is crucial. Hence, any delay caused by image processing is to be avoided. Furthermore, a fast algorithm like ours can even be imagined in an interventional setting.

While direct image-to-image mappings such as the method proposed by Ko et al. [40] circumvent the time-intensive optimization approach, they do not follow the two-step approach outlined in Fig. 1. Since they directly manipulate the image instead of estimating underlying motion curves explicitly, they risk a modification of the image content which is not consistent with the measured data. We observe this effect particularly for strong initial motion amplitudes which lead to altered anatomical structures that do not match the image content of the ground truth.

Concerning motion compensation performance, our method clearly outperforms the investigated existing methods. Total variation is a simple, non-parametric function which favors volumes with sparse gradients. However, it is agnostic to the content of the volume, the anatomy of the head in our case. Consequently, it can converge to solutions which represent implausible anatomy. Furthermore, TV might be more susceptible to the initialization and diverge for large initial motion patterns. Therefore, research has recently focused on learned quality metrics which are trained on examples of the target anatomy, thereby gaining an implicit understanding of it. The work by Huang et al. is an example for such method. We hypothesize that their method does not yield satisfactory results in our experiments because we fit a comparably large number of free parameters (180 compared to 72 free parameters in their original publication). This increases the complexity of the search space making the optimization task more difficult. Gradient-free optimization algorithms, in particular, exhibit unstable and slow convergence for large numbers of free parameters. We therefore think that the gradient-based optimization is not only advantageous in terms of speed, but also when increasing the number of free parameters.

In Fig. 4a, we observe that all methods have a higher MAE for both in-plane translations $t_x$ and $t_y$ than for the remaining motion parameters. This is, at least partly, due to the definition of the coordinate system with respect to the scanning trajectory. Both in-plane translations are parallel to the central ray for two positions in each full circular scan. The reconstructed image does not depend heavily on translations that are close to parallel to the central ray and, therefore, all investigated target functions are not as sensitive to these parameters as to the other motion parameters. However, since small errors in $t_x$ and $t_y$ barely influence the final reconstruction as long as they are roughly parallel to the measurement rays, we deem these deviations acceptable.

This study shows that our method performs motion com-pensation in a fast and robust way. Quantifying the effects on downstream task related to, e.g., stroke diagnosis or final patient outcome is crucial. First quantitative results have been published in prior work [9], but this aspect remains an open area for future research. We further acknowledge that motion artifacts are not the only source of image degradation in this context. Motion compensation always needs to go in hand with, e.g., scatter and beam hardening correction to obtain optimal reconstructed images. Further, the hardware of the CBCT scanner needs to be tailored to direct point-of-care imaging [4]. These are important aspects which are outside the scope of this work.

The proposed method is first tested on real MDCT reconstructions from which cone-beam sinograms are simulated. Realistic motion patterns are informed by existing research on the characteristics of head motion in an imaging device [8]. While simulated X-ray projections are known to have different characteristics compared to real ones [42], this difference largely vanishes during the reconstruction step because the reconstruction algorithm inverts the idealized image formation assumed during simulation. Furthermore, the constrained optimization setting of the proposed method might even tolerate slightly sub-optimal quality maps as long as they result in correct overall updates for the motion parameters. We prove this effect by applying our method to motion-affected clinical CBCT data. Importantly, we do not retrain the quality regression network. Nevertheless, motion artifacts are largely removed and the recovered motion patterns align with observable patterns extracted from the projection images proving the correctness of the recovered motion in the absence of a ground truth. These results indicate that the algorithm generalizes from the spline-based motion patterns to those induced by real patient motion. Furthermore, the slight differences in image appearance between reconstructions from direct CBCT measurements and simulations from MDCT data do not disturb the performance of the method. This is in contrast to the comparison methods by Ko et al., which produces strong artifacts in the caudal head region, and Huang et al., which does not generalize to the real data at all. A rigorous quantitative evaluation on the real CBCT data is challenging because ground-truth information about the patient's motion patterns are lacking. Hence, experiments on both data sets are required: The first setting uses partly simulated data for a large-scale and quantitative validation. The second part on motion-affected CBCT data demonstrates that results can be transferred to real data, but is limited in the number of available scans and quantitative evaluation. In follow-up work, it would be valuable to record CBCT scans along with their corresponding ground-truth motion patterns, e.g., using an optical tracking system. This can facilitate large-scale and quantitative studies of the performance across different variations of real motion patterns and their induced artifacts. Moreover, dedicated experiments with controlled levels of noise or scatter can yield a more comprehensive understanding of the method's performance when exposed to individual influencing variables.

The proposed algorithm fits a single rigid motion pattern to the entire reconstructed volume. As a result, multiple independently moving parts, such as head and stationary head rest or head and lower jaw, can not be modelled. Given the additive contribution of all voxels in the reconstructed volume to the gradient of the motion parameters, the largest region in volume domain would dominate the recovered motion signal. For brain examinations such as stroke imaging, the lower jaw is oftentimes not imaged and the head rest is clinically irrelevant such that some artifacts in this region are acceptable. Nevertheless, quantifying the influence of such independent rigid movements on a larger clinical data set is relevant for future research. Potentially, the presented method could be extended to multiple sets of projection matrices for different parts of the volume.

Gradient descent relies on accurate gradient information. We derive and implement an analytic expression to translate gradients on the reconstructed volume into gradients on the geometry parameters. In order to obtain precise geometry gradients, we require informative volume gradients which are obtained by standard backpropagation through the quality metric into the reconstructed volume. Since the quality metrics are usually trained on the forward mapping only, there is no guarantee that the gradient information from backpropagation is informative for motion compensation. For example, in a contracting architecture, the learned function could depend only on a subset of voxels, leading to missing gradient information for those positions that do not contribute to the output. We therefore regress full quality maps resulting in a quality metric prediction for every voxel in the volume. As a result, the learned mapping is enforced to incorporate all voxel positions and gradients flow back into all parts of the volume, which ensures more evenly distributed gradient information. In addition, skip connections in the U-net architecture are known to improve gradient flow and alleviate the problem of vanishing gradients [43]. Nevertheless, alternative volume-to-volume network architectures could be experimented with as well.

The analytic formulation used to propagate gradients from the reconstructed volume to the geometry parameters builds upon previously published work [26]. This prior work was restricted to fan-beam geometries which have limited relevance for clinical applications. Here, the idea is extended to cone-beam CT scanners. This extension requires a careful implementation of both backprojection and corresponding gradient computation to keep memory requirements manageable. While it is feasible to explicitly compute the entire Jacobian $\frac{d\mathbf{r}}{d\mathbf{p}}$ for fan-beam geometries, this would require approximately $290\,\text{GB}$ for a volume of size $256 \times 256 \times 256$ with 360 projection matrices. This is not compatible with existing consumer GPUs. As a result, the Jacobian is never computed explicitly in the cone-beam case, but Jacobian-vector products are evaluated on-the-fly using the gradient of the quality metric with respect to the reconstructed volume. Moreover, our previous work did not incorporate a motion model leading to unrealistically fast and temporally unconstrained simulated motion patterns which contradict existing knowledge about head motion characteristics.

Moreover, we would like to highlight that the geometry-differentiable implementation of the backprojection operation is openly available to the community as a ready-to-use *PyTorch* function which can be plugged into any existing autofocus-inspired approach to enable gradient-based optimization. It only requires a differentiable quality metric. Since most state-of-the-art approaches use neural networks for quality metric regression, differentiability is ensured through backpropagation. Consequently, gradient-based optimization is not only applicable in the specific context of this work, but has the potential to speed up existing autofocus-based works.

Some authors prefer gradient-free optimization algorithms for image alignment and motion compensation because these problems are typically not convex [19]. They believe gradient-based algorithms are more likely to terminate in sub-optimal local minima, while gradient-free methods can find more accurate solutions. However, both types of algorithms lack theoretical convergence guarantees for specific motion models and target functions. Moreover, our experiments consistently converge to solutions that are very close to the ground truth, both quantitatively and qualitatively. We acknowledge that gradient descent performs worse than CMAES for the TV quality metric, but no systematic difference can be identified between both optimization strategies for our proposed quality metric and the one proposed by Huang et al. If the combination of quality metric and motion model leads to a strong non-convex optimization landscape, we could imagine an approach which uses gradient descent for a fast initial alignment followed by fine-tuning of the estimated motion patterns with a gradient-free algorithm such as CMAES that might overcome some local minima due to its stochastic behavior. Moreover, in the deep learning community, research is conducted around gradient-based optimization algorithms which work robustly in non-convex optimization landscapes. For example, these algorithms use additional momentum terms or automatic parameter-dependent learning rate adaptation [44]. Replacing the plain gradient descent by a gradient-based optimization algorithm which is tailored toward non-convexity is also promising in strongly non-convex scenarios. Additionally, in non-convex optimization, the choice of the initial guess has a direct influence on the result. Since the estimation of motion patterns is patient- and scan-specific, there is usually no prior knowledge available and an initialization assuming a motion-free case seems logical. This is the strategy used in this work resulting in convincing results. Alternative ways of choosing the initial parameters are conceivable, e.g., a random spline-based motion pattern with small amplitude or an initial estimate by a separate neural network.

Another valuable direction for follow-up research is the extension of the proposed method to non-circular trajectories using analytic backprojection-type algorithms [45]. Not only would that allow for motion compensation for these types of trajectories, but a similar approach can be envisioned for phantom-free, image-based calibration for non-circular trajectories which can not be calibrated in advance.

## VI. CONCLUSION

In this paper, the differentiable formulation of CT reconstruction is extended to cone-beam geometry parameters making it applicable to motion compensation in real-world clinical scenarios. Together with an improved quality metric, we see the full advantage of the proposed gradient-based method: A substantial reduction of runtime as well as accurate and robust motion compensation performance. Ultimately, this method may pave the way toward point-of-care CBCT head imaging by correcting for inevitable motion artifacts in a fast and reliable manner.

## ACKNOWLEDGMENTS

The research leading to these results has received funding from the European Research Council (ERC) under the European Union's Horizon 2020 research and innovation program (ERC Grant No. 810316). The authors gratefully acknowledge the scientific support and HPC resources provided by the Erlangen National High Performance Computing Center of the Friedrich-Alexander-Universität Erlangen-Nürnberg. The hardware is funded by the German Research Foundation.

## DISCLAIMER

The concepts and information presented in this paper are based on research and are not commercially available.

## REFERENCES

[1] P. Nicholson *et al.*, "Novel flat-panel cone-beam CT compared to multi-detector CT for assessment of acute ischemic stroke: A prospective study," *Eur. J. Radiol.*, vol. 138, 5 2021.
[2] C. Waydhas, "Equipment review: Intrahospital transport of critically ill patients," *Crit. Care*, vol. 3, 1999.
[3] Z. Rumboldt, W. Huda, and J. W. All, "Review of portable CT with assessment of a dedicated head CT scanner," pp. 1630–1636, 10 2009.
[4] P. Wu *et al.*, "Cone-beam CT for imaging of the head/brain: Development and assessment of scanner prototype and reconstruction algorithms," *Med. Phys.*, vol. 47, pp. 2392–2407, 6 2020.
[5] M. Goyal *et al.*, "Analysis of workflow and time to treatment and the effects on outcome in endovascular treatment of acute ischemic stroke: Results from the SWIFT PRIME randomized controlled trial," *Radiology*, vol. 279, pp. 888–897, 6 2016.
[6] H. Huang *et al.*, "Reference-free learning-based similarity metric for motion compensation in cone-beam CT," *Phys. Med. Biol.*, vol. 67, no. 12, p. 125020, 2022.
[7] M. K. Kalra, M. M. Maher, R. D'Souza, and S. Saini, "Multidetector computed tomography technology: current status and emerging developments," *J. Comput. Assist. Tomogr.*, vol. 28, pp. S2–S6, 2004.
[8] A. Wagner, K. Schicho, F. Kainberger, W. Birkfellner, S. Grampp, and R. Ewers, "Quantification and Clinical Relevance of Head Motion during Computed Tomography," *Invest. Radiol.*, vol. 38, pp. 733–741, 2003.
[9] N. Cancelliere *et al.*, "Motion artifact correction for cone beam CT stroke imaging: a prospective series," *J. Neurointerv. Surg.*, vol. 15, pp. 223–228, 2023.
[10] M. Manhart *et al.*, "Denoising and artefact reduction in dynamic flat detector CT perfusion imaging using high speed acquisition: first experimental and clinical results," *Phys. Med. Biol.*, vol. 59, no. 16, p. 4505, 2014.
[11] H. Yu and G. Wang, "Data Consistency Based Rigid Motion Artifact Reduction in Fan-Beam CT," *IEEE TMI*, vol. 26, no. 2, pp. 249–260, 2007.
[12] M. Berger *et al.*, "Motion compensation for cone-beam CT using Fourier consistency conditions," *Phys. Med. Biol.*, vol. 62, no. 17, p. 7181, 2017.
[13] A. Aichert *et al.*, "Epipolar Consistency in Transmission Imaging," *IEEE TMI*, vol. 34, no. 11, pp. 2205–2219, 2015.
[14] A. Preuhs, A. Maier, M. Manhart, J. Fotouhi, N. Navab, and M. Unberath, "Double Your Views—Exploiting Symmetry in Transmission Imaging," in *Proc. MICCAI*. Springer, 2018, pp. 356–364.
[15] M. Berger *et al.*, "Marker-free motion correction in weight-bearing cone-beam CT of the knee joint," *Med. Phys.*, vol. 43, no. 3, pp. 1235–1248, 2016.

[16] S. Ouadah, J. Stayman, G. Gang, T. Ehtiati, and J. Siewerdsen, "Self-calibration of cone-beam CT geometry using 3D–2D image registration," *Phys. Med. Biol.*, vol. 61, no. 7, p. 2613, 2016.

[17] T. Sun, J.-H. Kim, R. Fulton, and J. Nuyts, "An iterative projection-based motion estimation and compensation scheme for head x-ray CT," *Medical physics*, vol. 43, no. 10, pp. 5705–5716, 2016.

[18] A. Kingston, A. Sakellariou, T. Varslot, G. Myers, and A. Sheppard, "Reliable automatic alignment of tomographic projection data by passive auto-focus," *Med. Phys.*, vol. 38, no. 9, pp. 4934–4945, 2011.

[19] A. Sisniega, J. W. Stayman, J. Yorkston, J. Siewerdsen, and W. Zbijewski, "Motion compensation in extremity cone-beam CT using a penalized image sharpness criterion," *Phys. Med. Biol.*, vol. 62, no. 9, p. 3712, 2017.

[20] A. Preuhs et al., "Appearance Learning for Image-Based Motion Estimation in Tomography," *IEEE TMI*, vol. 39, no. 11, pp. 3667–3678, 2020.

[21] S. Capostagno, A. Sisniega, J. Stayman, T. Ehtiati, C. Weiss, and J. Siewerdsen, "Deformable motion compensation for interventional cone-beam CT," *Phys. Med. Biol.*, vol. 66, no. 5, p. 055010, 2021.

[22] J. Wicklein, H. Kunze, W. A. Kalender, and Y. Kyriakou, "Image features for misalignment correction in medical flat-detector CT," *Med. Phys.*, vol. 39, no. 8, pp. 4918–4931, 2012.

[23] A. Sisniega et al., "Deformable image-based motion compensation for interventional cone-beam CT with a learned autofocus metric," in *Proc. SPIE*, vol. 11595, 2021, pp. 241–248.

[24] J. Wicklein, Y. Kyriakou, W. A. Kalender, and H. Kunze, "An online motion-and misalignment-correction method for medical flat-detector CT," in *Medical Imaging 2013: Physics of Medical Imaging*, vol. 8668. SPIE, 2013, pp. 466–472.

[25] H. Huang, J. H. Siewerdsen, A. Lu, Y. Hu, W. Zbijewski, M. Unberath, C. R. Weiss, and A. Sisniega, "Multi-stage Adaptive Spline Autofocus (MASA) with a learned metric for deformable motion compensation in interventional cone-beam CT," in *Medical Imaging 2023: Physics of Medical Imaging*, L. Yu, R. Fahrig, and J. M. Sabol, Eds., vol. 12463, International Society for Optics and Photonics. SPIE, 2023, p. 1246314. [Online]. Available: https://doi.org/10.1117/12.2654361

[26] M. Thies et al., "Gradient-based geometry learning for fan-beam CT reconstruction," *Phys. Med. Biol.*, vol. 68, no. 20, p. 205004, 2023.

[27] M. Thies et al., "Optimizing CT Scan Geometries With and Without Gradients," in *Proc. Fully3D*, 2023.

[28] H. Huang, J. H. Siewerdsen, W. Zbijewski, C. R. Weiss, M. Unberath, and A. Sisniega, "Context-aware, reference-free local motion metric for CBCT deformable motion compensation," in *7th International Conference on Image Formation in X-Ray Computed Tomography*, J. W. Stayman, Ed., vol. 12304, International Society for Optics and Photonics. SPIE, 2022, p. 1230412. [Online]. Available: https://doi.org/10.1117/12.2646857

[29] X. Li, D. Zhang, and B. Liu, "A generic geometric calibration method for tomographic imaging systems with flat-panel detectors—A detailed implementation guide," *Med. Phys.*, vol. 37, no. 7, pp. 3844–3854, 2010.

[30] A. Maier, S. Steidl, V. Christlein, and J. Hornegger, *Medical imaging systems: An introductory guide*. Springer, 2018.

[31] J. Maier et al., "Rigid and Non-Rigid Motion Compensation in Weight-Bearing CBCT of the Knee Using Simulated Inertial Measurements," *IEEE TBE*, vol. 69, no. 5, pp. 1608–1619, 2021.

[32] C. Rohkohl, G. Lauritsch, L. Biller, M. Prümmer, J. Boese, and J. Hornegger, "Interventional 4D motion estimation and reconstruction of cardiac vasculature without motion periodicity assumption," *Med. Image Anal.*, vol. 14, no. 5, pp. 687–694, 2010.

[33] H. Wu, A. Maier, R. Fahrig, and J. Hornegger, "Spatial-temporal total variation regularization (STTVR) for 4D-CT reconstruction," in *Proc. SPIE Medical Imaging*, vol. 8313. SPIE, 2012, pp. 1018–1024.

[34] H. R. Sheikh and A. C. Bovik, "Image information and visual quality," *IEEE TIP*, vol. 15, pp. 430–444, 2 2006.

[35] Y. Shao, F. Sun, H. Li, and Y. Liu, "A novel approach for computing quality map of visual information fidelity index," in *Proc ISKE*, vol. 213. Springer, 2014, pp. 163–173.

[36] A. Wolny et al., "Accurate and versatile 3d segmentation of plant tissues at cellular resolution," *eLife*, vol. 9, p. e57613, jul 2020.

[37] V. Bacher, C. Syben, A. Maier, and A. Wang, "Learning projection matrices for marker free motion compensation in weight-bearing CT scans," in *Proc. Fully 3D*, vol. 16, 2021, pp. 327–330.

[38] S. Chilamkurthy et al., "Deep learning algorithms for detection of critical findings in head CT scans: a retrospective study," *The Lancet*, vol. 392, no. 10162, pp. 2388–2396, 2018.

[39] C. Syben, M. Michen, B. Stimpel, S. Seitz, S. Ploner, and A. K. Maier, "PYRO-NN: Python reconstruction operators in neural networks," *Med. Phys.*, vol. 46, no. 11, pp. 5110–5115, 2019.

[40] Y. Ko, S. Moon, J. Baek, and H. Shim, "Rigid and non-rigid motion artifact reduction in X-ray CT using attention module," *Medical Image Analysis*, vol. 67, p. 101883, 2021.

[41] N. Hansen, "The CMA evolution strategy: a comparing review," in *Towards a New Evolutionary Computation*, ser. Studies in Fuzziness and Soft Computing. Springer, 2006, vol. 192, pp. 75–102.

[42] C. Gao et al., "Synthetic data accelerates the development of generalizable learning-based algorithms for X-ray image analysis," *Nat. Mach. Intell.*, vol. 5, no. 3, pp. 294–308, 2023.

[43] M. Drozdzal, E. Vorontsov, G. Chartrand, S. Kadoury, and C. Pal, "The importance of skip connections in biomedical image segmentation," in *Proc. DLMIA, LABELS*. Springer, 2016, pp. 179–187.

[44] D. P. Kingma and J. Ba, "Adam: A method for stochastic optimization," *arXiv preprint arXiv:1412.6980*, 2014.

[45] M. Defrise and R. Clack, "A cone-beam reconstruction algorithm using shift-variant filtering and cone-beam backprojection," *IEEE TMI*, vol. 13, no. 1, pp. 186–195, 1994.